# 2-DOF BLOCK POLE PLACEMENT CONTROL APPLICATION TO: HAVE-DASH-II BTT MISSILE

BEKHITI Belkacem[1] DAHIMENE Abdelhakim[1] NAIL Bachir[2] and HARICHE Kamel[1]

[1]Electronics and Electrotechnics Institute, University of Boumerdes, 35000 Algeria.
Belkacem1988@hotmail.co.uk
[2] Technology and sciences Institute, University of Djelfa, Algeria
nail_bachir@yahoo.fr

## ABSTRACT

*In a multivariable servomechanism design, it is required that the output vector tracks a certain reference vector while satisfying some desired transient specifications, for this purpose a 2DOF control law consisting of state feedback gain and feedforward scaling gain is proposed. The control law is designed using block pole placement technique by assigning a set of desired Block poles in different canonical forms. The resulting control is simulated for linearized model of the HAVE DASH II BTT missile; numerical results are analyzed and compared in terms of transient response, gain magnitude, performance robustness, stability robustness and tracking. The suitable structure for this case study is then selected.*

## 1. INTRODUCTION

Unlike other engineering specialties whose subject of study is a specific engineering system such as an engine system or an airborne system, control systems theory studies only a general mathematical model of engineering systems. The reason that systems control theory has concentrated mainly on linear time-invariant systems is that only the mathematical models of this kind of systems can have general and explicit solutions. Furthermore, only the general and explicit understanding of the system can be used to guide generally, systematically, and effectively the complicated control system design [1]. The purpose and requirement of control systems is generally the control of plant system output (or response) Y(s) so that it can quickly reach and stabilize to its desired state, such as the desired vehicle and engine speed, the desired radar and airborne system angle, the desired robot arm position, the desired container pressure and temperature, etc. The desired system output state is usually specified by the reference signal R(s). Hence how well the system output reaches its desired state determines the performance of the system; see [1, 4 and 17]. The control problems associated with these systems might be the production of some chemical product as efficiently as possible, automatic landing of aircraft, rendezvous with an artificial satellite, regulation of body functions such as heartbeat or blood pressure, and the ever-present problem of economic inflation [18]. State space control theory provides distinctly general, accurate, and clear analysis on linear time-invariant systems, especially their performance and sensitivity properties. Only this kind of analysis and understanding can be used to guide generally and effectively the design of complex control systems. This is the reason that linear time-invariant system control results form the basis of the study of other systems such as nonlinear, distributive, and time-varying systems, even though most practical systems belong to the latter category. This is also the reason that the development of state space control theory has always been significant and useful. [17]. A large-scale MIMO system, described by state equations, is often decomposed into small subsystems, from which the analysis and design of the MIMO system can be easily performed. Similarity block transformations are developed to transform a class of linear time-invariant MIMO state equations, for which the systems described by these equations have the number of inputs dividing exactly the order of the state, into block companion forms so that the classical lines of thought for SISO systems can be extended to MIMO systems [19]. Such systems can be studied via the eigenstructure, eigenvalues and eigenvectors, of the state matrix A. The eigenvalues and eigenvectors can determine system performance and robustness far more directly and explicitly than other indicators. Hence their assignment should improve feedback system performance and robustness distinctly and effectively [1]. Eigenstructure assignment (EA) is the process of applying negative feedback to a linear, time-invariant system with the objective of forcing the latent-values and latent-vectors to become as close as possible to a desired eigenstructure. EA, in common with other multivariable design methodologies, is inclined to use all of the available design freedom to generate a control solution. It is a natural choice for the design of any control system whose desired performance is readily represented in terms of an ideal eigenstructure. Many research works has been done on EA [20, 21, 22, 23 and 24] and more specifically on flight control systems [25, 26, and 27].

The critical importance of system poles (eigenvalues of system dynamic matrix) on system performance are determined and examined by the location of those roots, however in a complementary the sensitivity of eigenvalues is determined by their corresponding eigenvectors which is a basic result of numerical linear algebra. Unfortunately numerical linear algebra, has not been commonly used in the existing textbooks on control systems, is a branch of study which concentrates on the sensitivity of linear algebraic computation with respect to the initial data variation and computational round-off errors [Fox, 1964]. Because linear algebra is the basic mathematical tool in linear control systems theory, the results of numerical linear algebra can be used directly in analyzing linear system sensitivities. When the controlled system is multi-input multi-output then an infinite number of gain matrices K may be found which will provide the required stability characteristics. Consequently, an alternative and very powerful method for designing feedback gains for auto-stabilization systems is the right and/or left block pole placement method. The method is based on the manipulation of the equations of motion in block state space form and makes full use of the appropriate computational tools in the analytical process. The design of state feedback control in MIMO systems leads to the so-called matrix polynomials assignment [2]. The use of block poles constructed from a desired set of closed-loop poles offers the advantage of assigning a characteristic matrix polynomial rather than a scalar one [3]. The desired characteristic matrix polynomial is first constructed from a set of block poles selected among a class of similar matrices, and then the state feedback is synthesized by solving matrix equations. The forms of the block poles used in our work are the diagonal, the controller and the observer forms. Robustness is assessed, in each case, using the infinity norm, the singular value of the closed loop transfer matrix and the condition number of the closed-loop transfer matrix. Time response is assessed by plotting the step response and comparing the time response characteristics [1]. A comparison study is conducted to determine, in light of the above criteria, the best choice of the form of the block poles.

In the present paper, firstly we have started the work by introducing some theoretical preliminaries on matrix polynomials, after that a theoretical background on robustness and sensitivity analysis in term of responses is illustrated and briefly discussed, it is then followed by an application to BTT missiles by doing a comparison study in term of block roots form. As a fifth section a discussion of the obtained results is performed, and finally the paper is finished by a comparison study and a conclusion.

## 2. PRELIMINARIES

### 2.1. Definition of a polynomial matrix

**Definition:** given a set of $m \times m$ complex matrices $\{A_0, A_1, \ldots, A_l\}$ the following matrix valued function of the complex variable $\lambda$ is called matrix polynomial of degree (index) $l$ and order $m$ : ($A(\lambda)$: is called also $\lambda$-matrix.)

$$A(\lambda) = A_0 \lambda^l + A_1 \lambda^{l-1} + \cdots + A_{l-1} \lambda + A_l \quad (1)$$

Consider the system described by the following dynamic equation:

$$\begin{cases} \dot{x}(t) = Ax(t) + Bu(t) \\ y(t) = Cx(t) \end{cases}, \quad \text{With } A \in R^{n \times n}, B \in R^{n \times m}, C \in R^{p \times n} \quad (2)$$

Assuming that the system can be transformed to a block controller form, this means:
i. The number $\frac{n}{m} = l$ is an integer.
ii. The matrix $W_c = \{B, AB, \ldots, A^{l-1}B\}$ is of full rank n.

Then we use the following transformation matrix:

$$T_c = \begin{bmatrix} T_{cl} \\ T_{cl}A \\ \vdots \\ T_{cl}A^{l-2} \\ T_{cl}A^{l-1} \end{bmatrix} \quad \text{Where } T_{cl} = [O_m \; O_m, \ldots, I_m][B, AB, \ldots, A^{l-1}B]^{-1} \quad (3)$$

The new system becomes:

$$\begin{cases} \dot{x}(t) = A_c x(t) + B_c u(t) \\ y(t) = C_c x(t) \end{cases} \quad (4)$$

With

$$A_c = T_c A T_c^{-1}, \quad B_c = T_c B, \quad C_c = C T_c^{-1}$$

$$A_c = \begin{bmatrix} O_m & I_m & \cdots & O_m \\ O_m & O_m & \cdots & O_m \\ \vdots & \vdots & \cdots & \vdots \\ O_m & O_m & \cdots & I_m \\ -A_l & -A_{l-1} & \cdots & -A_1 \end{bmatrix}, \quad B_c = \begin{bmatrix} O_m \\ O_m \\ \vdots \\ O_m \\ I_m \end{bmatrix}, \quad C_c = [C_l \; C_{l-1} \; \cdots \; C_1]$$

## 2.2. Matrix transfer function

The matrix transfer function of this open-loop system is given by:

$$TF_R(s) = G(s) = N_R(s)D_R^{-1}(s) \quad (5)$$

Where:

- $N_R(s) = [C_1 s^{l-1} + \cdots + C_{l-1} s + C_l]$
- $D_R(s) = [I_m s^l + A_1 s^{l-1} + \cdots + A_l]$

This transfer function is called the Right Matrix Fraction Description (RMFD); we need to use it in the block controller form. It should be noted that the behavior of the system depends on the characteristic matrix polynomial $D_R(s)$.

## 2.3. Concept of solvents (block roots)

A root for a polynomial matrix is not well defined. If it is defined as a complex number it may not exist at all. Then we may consider a root as a matrix called block root.

### 2.3.1. Right solvent

Given the matrix polynomial of order $m$ and index $l$ defined by:

$$D_R(s) = I_m s^l + A_1 s^{l-1} + \cdots + A_l = \sum_{i=0}^{l} A_i s^{l-i} \text{ where } A_0 = I_m \quad (6)$$

A right solvent, denoted by $R$, is a $m \times m$ matrix satisfying:

$$D_R(R) = A_0 R^l + A_1 R^{l-1} + \cdots + A_{l-1} R + A_l = 0_m \quad (7)$$

### 2.3.2. Left solvent

A left solvent of the matrix polynomial $D(s)$ defined above, denoted by $L$, is $m \times m$ matrix satisfying: $\quad D_R(L) = L^l A_0 + L^{l-1} A_1 + \cdots + L A_{l-1} + A_l = 0_m \quad (8)$

A right solvent, if exist, is considered as a right block root. A left solvent, if exist, is considered as a left block root.

### 2.3.3. Latent root and latent vector

- A complex number $\lambda$ satisfying $\det(D_R(\lambda)) = 0$ is called a latent root of $D_R(\lambda)$.
- Any vector $x_i$ associated with the latent root satisfying $D_R(\lambda_i) x_i = 0_m$ is a right latent vector of $D_R(\lambda)$.

The relationship between latent roots, latent vectors, and the solvents can be stated as follows:

**Theorem:** If $D(\lambda)$ has $n$ linearly independent right latent vectors $p_1, p_2, \ldots, p_n$ (left latent vectors $q_1, q_2, \ldots, q_n$) corresponding to latent roots $p_1, p_2, \ldots, p_n$, then $P \Lambda P^{-1}$, $(Q \Lambda Q^{-1})$ is a right (left) solvent. Where: $P = [p_1, p_2, \ldots, p_n]$, $(Q = [q_1, q_2, \ldots, q_n]^T)$ and $\Lambda = diag(\lambda_1, \lambda_2, \ldots, \lambda_n)$. **Proof:** see [16]

**Theorem:** If $D(\lambda)$ has $n$ latent roots $\lambda_1, \lambda_2, \ldots, \lambda_n$ and the corresponding right latent vectors $p_1, p_2, \ldots, p_n$ has as well as the left latent vectors $q_1, q_2, \ldots, q_n$ are both linearly independent, then the associated right solvent $R$ and left solvent $L$ are related by: $R = WLW^{-1}$, Where $W = PQ$ and $P = [p_1, p_2, \ldots, p_n]$, ($Q = [q_1, q_2, \ldots, q_n]^T$). and "$T$" stands for transpose. (**Proof:** see [16] )

### 2.3.4. Complete set of solvents

**Definition :** Consider the set of solvents $\{R_1, R_2, \ldots, R_l\}$ constructed from the eigenvalues $(\lambda_1, \lambda_2, \ldots, \lambda_n)$ of a matrix $A_c$, $\{R_1, R_2, \ldots, R_l\}$ is a complete set of solvents if and only if:

$$\begin{cases} \cup \sigma(R_i) = \sigma(A_c) \\ \sigma(R_i) \cap \sigma(R_j) = \emptyset \\ \det(V_R(R_1, R_2, \ldots, R_l)) \neq 0 \end{cases} \quad (9)$$

Where:
$\sigma$ denotes the spectrum of the matrix.
$V_R$ is the block Vandermonde matrix corresponding to $\{R_1, R_2, \ldots, R_l\}$ given as:

$$V_R(R_1, R_2, \ldots, R_l) = \begin{bmatrix} I_m & I_m & \cdots & I_m \\ R_1 & R_2 & \cdots & R_l \\ \vdots & \vdots & \ddots & \vdots \\ R_1^{l-1} & R_2^{l-1} & \cdots & R_l^{l-1} \end{bmatrix} \quad (10)$$

The conditions for the existence and uniqueness of the complete set of solvents have been investigated by P. Lancaster [15] and Malika Yaici [3].

**Remark:** We can define a set of left solvents in the same way as in the previous theorem.

### 2.4. Constructing a matrix polynomial from a complete set of solvents

We want to construct the matrix polynomial defined by $D(\lambda)$ from a set of solvents or a set of desired poles which will determine the behavior of the system that we want. Suppose we have a desired complete set of solvents. The problem is to find the desired polynomial matrix or the characteristic equation of the block controller form defined by:

$$D(\lambda) = D_0 \lambda^l + D_1 \lambda^{l-1} + \cdots + D_l$$

We want to find the coefficients $D_i$ for $i = 1, \ldots, l$

**a. Constructing from a complete set of right solvents:**

Consider a complete set of right solvents $\{R_1, R_2, \ldots, R_l\}$ for the matrix polynomial $D(\lambda)$, If $R_i$ is a right solvent of $D(\lambda)$ so: $R_i^l + D_1 R_i^{l-1} + \cdots + D_l = 0_m \Rightarrow D_1 R_i^{l-1} + \cdots + D_l = -R_i^l$

Replacing $i$ from $1\ to\ l$ we get the following:

$$[D_{dl}, D_{d(l-1)}, \ldots, D_{d1}] = -[R_1^l, R_2^l, \ldots, R_l^l] V_R^{-1} \quad (11)$$

Where $V_R$ is the right block Vandermonde matrix

### b. Constructing from a complete set of left solvents:

Consider a complete set of left solvents $\{L_1, L_2, \ldots, L_l\}$ for the matrix polynomial $D(\lambda)$ If $L_i$ is a left solvent of $D(\lambda)$ so: $L_i^l + L_i^{l-1}D_1 + \cdots + D_l = O_m \Rightarrow L_i^{l-1}D_1 + \cdots + D_l = -L_i^l$

Replacing $i$ from 1 to $l$ we get the following:

$$\begin{bmatrix} D_{dl} \\ D_{d(l-1)} \\ \vdots \\ D_{d1} \end{bmatrix} = - \begin{bmatrix} I_m & L_1 & \cdots & L_1^{l-1} \\ I_m & L_2 & \cdots & L_2^{l-1} \\ \vdots & \vdots & \ddots & \vdots \\ I_m & L_l & \cdots & L_l^{l-1} \end{bmatrix}^{-1} \begin{bmatrix} L_1^l \\ L_2^l \\ \vdots \\ L_l^l \end{bmatrix} \quad (12)$$

### 2.5. State feedback design

Consider the general linear time-invariant dynamic system described by the previous state space equation (2). Now applying the state feedback $u = -K_{FB}x(t)$ to this system, where: $K_{FB}$ is a $m \times n$ gain matrix. After using the block controller form transformation for the system, we get: $u = -K_c x_c(t)$

Where: $K_{FB} = K_c T_c = [K_{cl}, K_{c(l-1)}, \ldots, K_{c1}]T_c$ and $K_{ci} \in R^{m \times m}$ for $i = 1, \ldots, l$. Then the resulting closed loop system is shown below:

$$\begin{cases} \dot{x}_c = (A_c - B_c K_c)x_c \\ y_c = C_c x_c \end{cases} \text{Where: } (A_c - B_c K_c) = \begin{bmatrix} O_m & \cdots & O_m \\ O_m & \cdots & O_m \\ \vdots & \cdots & \vdots \\ O_m & \cdots & I_m \\ -(A_l + K_{cl}) & \cdots & -(A_1 + K_{c1}) \end{bmatrix}$$

The characteristic matrix polynomial of this closed loop system is:

$$D(\lambda) = I_m \lambda^l + (A_1 + K_{c1})\lambda^{l-1} + \cdots + (A_l + K_{cl}) \quad (13)$$

From a set of desired eigenvalues, we construct the solvents then we construct the desired characteristic matrix polynomial in the form:

$$D_d(\lambda) = I_m \lambda^l + D_{d1}\lambda^{l-1} + \cdots + D_{dl} \quad (14)$$

By putting $D_d(\lambda) = D(\lambda)$ we get the coefficients $K_{ci}$ as follows:

$$K_{ci} = D_{di} - A_i \quad \text{for } i = 1, \ldots, l \quad (15)$$

After that we find the gain matrix by the following formula $K_{FB} = K_c T_c$.

### 2.6. Feedforward gain design

In this subsection the feedforward gain will be determined in order to provide steady state tracking. Consider the MIMO system of equation (2), the control law is then given by $u(t) = -K_{FF}r(t) - K_{FB}x(t)$ Where: $K_{FB}$ is the feedback gain matrix obtained one of the techniques discussed in the previous section, $K_{FF}$ is the feedforward gain in question and $r(t) = r$ is a constant vector. The steady state is defined by the next conditions:

$$x(\infty) = \lim_{t \to \infty} x(t) = \text{Constant}, \quad \dot{x}(\infty) = 0, \quad y(\infty) = r$$

The feedforward $K_{FF}$ gain must be chosen so that the reference $r$ lies within the null space of: $Q = [I - (C - DK_{FB})(A - BK_{FB})^{-1}BK_{FB} + DK_{FF}]$. A trivial solution is to choose $Q = 0$ so that every vector $r$ satisfies the condition.

$$K_{FF} = [(C - DK_{FB})(A - BK_{FB})^{-1}B + D]^+ \tag{16}$$

**Remark:** The notation $(+)$ denotes the pseudo-inverse; and this method does not guarantee the existence of solution in the case where: $[(C - DK_{FB})(A - BK_{FB})^{-1}B + D]$ is not full rank. Strictly speaking, its existence is governed by the following theorem.

**Theorem:** The feedforward gain that provides steady state decoupling and tracking exist if and only if: $\operatorname{rank}\left(\begin{bmatrix} A & B \\ C & D \end{bmatrix}\right) = n + m$ (**Proof:** see [30] )

**Remark:** The feedforward gain depends on system matrices and the feedback gain $K_{FB}$, hence it is influenced by perturbation, modeling error and the feedback gain $K_{FB}$.

## 3. ROBUSTNESS AND SENSITIVITY ANALYSIS

One of the major concerns in control design is system's immunity to modeling errors and different types of disturbances that may affect it, this is known as robustness. This issue will be addressed in this section.

### 3.1. Basic definitions of matrix norms

**Definition:** A matrix norm is a function from the set of all complex matrices (of all finite orders) into that satisfies the following properties [1], [28]:

1. $\|A\| \geq 0$ and $\|A\| \geq 0 \Rightarrow A = 0$  　　2. $\|\alpha A\| = \alpha \|A\|$　　for all scalars $\alpha$
3. $\|A + B\| \leq \|A\| + \|B\|$　for all matrices of the same size.
4. $\|AB\| \geq \|A\|\|B\|$　for all conformable matrices.

**Definition:** The most commonly used norms are the following

1. The matrix 1-Norm is defined as the largest absolute column sum, given by
$$\|A\|_1 = \max_j \sum_i |a_{ij}| \tag{17}$$

2. The matrix 2-Norm is defined as the maximum singular value of $A$, given by
$$\|A\|_2 = \text{Max}(\text{singular value of } A) = \text{Max}\left(\text{eig}(A^\star A)^{\frac{1}{2}}\right) \tag{18}$$

3. The matrix ∞-Norm is defined as the largest absolute row sum, given by
$$\|A\|_\infty = \max_i \sum_i |a_{ij}| \tag{19}$$

4. The Forbenius Norm is also called Shure norm, it is defined as a square root of the trace $A^\star A$ given by
$$\|A\|_F = \text{Trace}(A^\star A)^{\frac{1}{2}} = \left(\sum_{i,j} |a_{ij}|^2\right)^{\frac{1}{2}} \tag{20}$$

## 3.2. The sensitivity of eigenvalues (robust performance)

Robust performance is defined as the low sensitivity of system performance with respect to system model uncertainty and terminal disturbance. It is well known that the eigenvalues of the dynamic matrix determine the performance of the system then from that the sensitivities of these eigenvalues determine the robustness of the system (2).

**Theorem:** Let $\lambda$ and $\lambda'$ be the eigenvalues of the matrices $A$ and $A + \Delta A$ respectively, and let $V$ be the right eigenvectors matrix of $A$, then Wilkinson has derived the variation in eigenvalues as follows: $\min_i(\lambda_i - \lambda_i') \triangleq \min_i(\Delta(\lambda_i)) \leq \kappa(V).\|\Delta A\|$ (21)
$\|.\|$ Stands for the matrix norm and $\kappa(.)$ Is the condition number. (**Proof:** see [1])

**Theorem:** Let $\lambda_i, \boldsymbol{v_i}$ and $\boldsymbol{t_i}$ be the $i^{th}$ eigenvalue, right and left eigenvectors of a matrix $A$ respectively $(i = 1, \dots, n)$, let $\lambda_i + \Delta \lambda_i$ be the $i^{th}$ eigenvalue of the matrix $A + \Delta A$, then for small enough $\|\Delta A\|$: $\quad \Delta \lambda_i \leq \|\boldsymbol{v_i}\|\|\boldsymbol{t_i}\|\|\Delta A\| \triangleq s(\lambda_i)\|\Delta A\|$ (22)
Such that: $s(\lambda_i) = \|\boldsymbol{v_i}\|\|\boldsymbol{t_i}\|$. (**Proof:** see [1].)

This theorem shows that the sensitivity of an eigenvalue is determined by its corresponding left and right eigenvectors and it is valid for small perturbations in the matrix $A$.

## 3.3. Relative change

Let $\lambda_i$ and $\lambda_i'$ be the eigenvalues of the matrices $A$ and $A + \Delta A$ respectively. The relative change $r_i$ of the eigenvalue $\lambda_i$ is defined as follows:

$$r_i = \frac{|\lambda_i - \lambda_i'|}{|\lambda_i|} = \frac{|\Delta \lambda_i|}{|\lambda_i|} \quad i = 1,\dots,n \quad (23)$$

## 3.4. Robust stability

The stability of a system is the most wanted property. So its sensitivity to uncertainties is very important when analyzing and designing the system. Stability is affected by the system eigenvalues of the dynamic matrix so the sensitivity of these eigenvalues directly affects the robust stability of the system (2). There are three robust stability measures using the sensitivity of this system eigenvalues defined as follows:

**Definition:** Let $\{\lambda_1, \lambda_2, \dots, \lambda_n\}$ be the set of eigenvalues of an $n \times n$ matrix denoted by $A$ and assuming that all the eigenvalues are stable $(i.e: Re\{\lambda_i\} < 0 \; \forall i)$ and all the eigenvalues are already arbitrary assigned for guaranteed performance, the three robust stability measures are defined by:

1. $M_1 = \min_{0<\omega<\infty}\{\underline{\sigma}(A - j\omega I)\}$, $\underline{\sigma}$ denotes the smallest singular value.
2. $M_2 = (\kappa(\Lambda))^{-1}.|Re\{\lambda_n\}|$ such that: $|Re\{\lambda_n\}| \leq \cdots \leq |Re\{\lambda_1\}|$ and $\Lambda$ is the diagonal matrix of $A$.
3. $M_3 = \min_{0 \leq i \leq n}\{(s(\lambda_i))^{-1}|Re\{\lambda_i\}|\}$

## 3.5. Tracking robustness

Consider the MIMO system with reference vector $r$ and output $y$ which is controlled by the control law under study, beside that the modeling errors and disturbances can be modeled by a perturbation matrix $\Delta A$, the tracking error induced by this perturbation in the S-domain is:

$$E(s) = R(s) - Y(s) = R(s) - K_{FF}G(s)R(s) \tag{24}$$

Or in more compact form

$$E(s) = [I + K_{FF}C[sI - (A + \Delta A - BK_{FB})]^{-1}B]R(s) \tag{25}$$

To find the steady state error vector the final value theorem is used:

$$e(\infty) = \lim_{s \to 0} sE(s) = [I - K_{FF}C(A + \Delta A - BK_{FB})^{-1}B]r \tag{26}$$

For perturbation small enough we have [28]:

$$(A + \Delta A - BK_{FB})^{-1} \approx (A - BK_{FB})^{-1} - (A - BK_{FB})^{-1}\Delta A(A - BK_{FB})^{-1} \tag{27}$$

Equation (26) become then: $e(\infty) = [K_{FF}C(A - BK_{FB})^{-1}\Delta A(A - BK_{FB})^{-1}B]r$. Assume for simplicity $D = 0$, hence replacing by its value gives:

$$e(\infty) = [B^+\Delta A(A - BK_{FB})^{-1}B]r \tag{28}$$

Applying the 2-norm on this equation yields:

$$\frac{\|e(\infty)\|_2}{\|r\|_2} \leq \|B^+\|_2 \|B\|_2 \|(A - BK_{FB})^{-1}\|_2 \|\Delta A\|_2 \tag{29}$$

**Discussion:** from this equation it can be seen clearly that the relative tracking error due to the perturbation depends on the closed loop matrix, hence we expect it to be different for various state feedback schemes.

## 4. DYNAMIC MODELING AND CONTROL OF BTT MISSILE

A missile is defined as a space-traversing unmanned vehicle the means for controlling its movements and estimating its flight path. Missiles can be classified according to their area of launching and the target's area into the following four categories: ground to ground, ground to air, air to air and air to ground, in this work the air to ground bank-to-turn (BTT) missiles will be considered. For the purpose of this study, only the control function is considered. In the following sections the 6DOF dynamic model of the missile is derived which is then linearized about the steady state conditions in order to obtain a linear state space model of the missile suitable for study. Afterward the controller objectives are stated and the output equations are derived with design requirement and specifications.

## 4.1. Missile Dynamics

Define the following reference frames and fundamental missile movements:

- Missile body fixed reference frame $(X, Y, Z)$ with its origin at the center of gravity of the missile and axis point to the missiles nose (Figure 1).
- The space fixed (nonrotating) reference frame $(\bar{X}, \bar{Y}, \bar{Z})$ with its origin at the center of gravity of the missile, which by performing Euler rotation $(\phi, \theta, \psi)$ corresponding to the roll, pitch and yaw respectively come to coincide with the $(X, Y, Z)$ reference frame see (Figure 1).
- Translation along the $V$ direction (velocity) denoted $V = (u, v, w)^T$.
- Rotation about the longitudinal axis (roll) denoted $p$.
- Rotation about the lateral horizontal axis (pitch) denoted $q$.
- Rotation about the vertical axis (yaw) denoted $r$. and $\omega = (p, q, r)^T$ be the angular velocity of missile with respect to $(\bar{X}, \bar{Y}, \bar{Z})$ fixe reference frame.

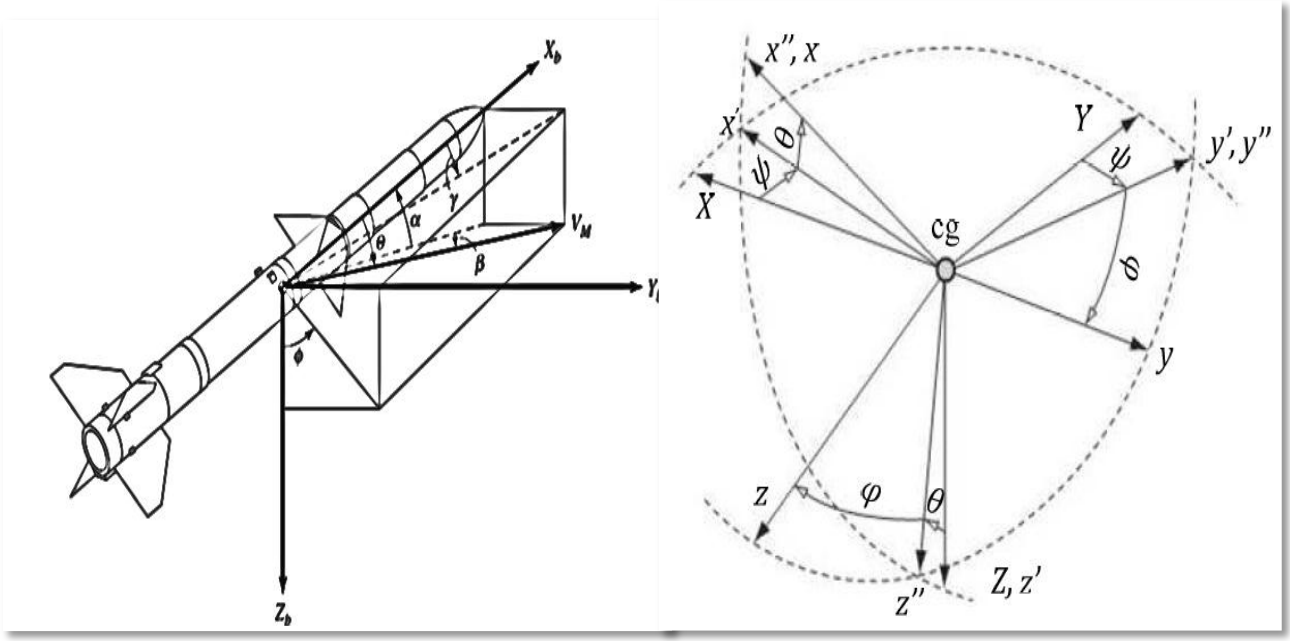

Figure: 1 Missile configuration and Euler angles.

**Assumptions:** The missile dynamic equation derived under the following assumptions:

- The missile is a rigid body.
- The mass $m$ and inertia $I$ of the missile remain constant over the period of time autopilot operation.
- The missile presents both mass and geometry symmetry with respect to $XZ$ and $XY$ planes, which means that the coupling inertia terms $I_{ij}$ for $i, j = x, y, z$ and $i \neq j$ are zeros.

The nonlinear dynamic model of the BTT missile is described by the next set of coupled differential equations [29]:

$$\begin{cases} \dot{\alpha} = q - p\beta + \left(\dfrac{1}{mV_m}\right)\left[k_f \rho V_m^2 (C_{z0} + C_{ze}\delta_e + C_{za}\delta_a + C_{zr}\delta_r) + mg.\cos(\phi)\right] \\ \dot{\beta} = -r + p\alpha + \left(\dfrac{1}{mV_m}\right)\left[k_f \rho V_m^2 (C_{y0} + C_{ye}\delta_e + C_{ya}\delta_a + C_{yr}\delta_r) + mg.\sin(\phi)\right] \\ \dot{\phi} = p \\ \dot{p} = \left(\dfrac{I_{yy} - I_{zz}}{I_{xx}}\right) qr + \left(\dfrac{1}{I_{xx}}\right)\left[k_m \rho V_m^2 (C_{l0} + C_{le}\delta_e + C_{la}\delta_a + C_{lr}\delta_r)\right] \\ \dot{q} = \left(\dfrac{I_{zz} - I_{xx}}{I_{yy}}\right) pr + \left(\dfrac{1}{I_{yy}}\right)\left[k_m \rho V_m^2 (C_{m0} + C_{me}\delta_e + C_{ma}\delta_a + C_{mr}\delta_r)\right] \\ \dot{r} = \left(\dfrac{I_{xx} - I_{yy}}{I_{zz}}\right) pq + \left(\dfrac{1}{I_{zz}}\right)\left[k_m \rho V_m^2 (C_{n0} + C_{ne}\delta_e + C_{na}\delta_a + C_{nr}\delta_r)\right] \end{cases}$$

Where: $\rho$ is the atmospheric density, $k_f$ and $k_m$ are constant determined by the vehicle geometry. $\boldsymbol{C_{F0}}, \boldsymbol{C_{Fu}}, \boldsymbol{C_{M0}}$ and $\boldsymbol{C_{Mu}}$ are the missile's aerodynamic coefficients given in below:

$$\boldsymbol{C_{F0}} = \begin{bmatrix} C_{x0} \\ C_{y0} \\ C_{z0} \end{bmatrix} = \begin{bmatrix} -0.57 + 0.083\alpha \\ -0.21\beta \\ C_{z0}(\alpha, M_m) \end{bmatrix}, \quad \boldsymbol{C_{Fu}} = \begin{bmatrix} C_{xe} & C_{xa} & C_{xr} \\ C_{ye} & C_{ya} & C_{yr} \\ C_{ze} & C_{za} & C_{zr} \end{bmatrix} = \begin{bmatrix} 0.04 & 0.00 & 0.00 \\ 0.00 & 0.00 & 0.08 \\ -0.09 & 0.00 & 0.00 \end{bmatrix}$$

$$\boldsymbol{C_{M0}} = \begin{bmatrix} C_{l0} \\ C_{m0} \\ C_{n0} \end{bmatrix} = \begin{bmatrix} -0.116\beta \\ C_{m0}(\alpha, M_m) \\ 0.08 \end{bmatrix}, \quad \boldsymbol{C_{Mu}} = \begin{bmatrix} C_{le} & C_{la} & C_{lr} \\ C_{me} & C_{ma} & C_{mr} \\ C_{ne} & C_{na} & C_{nr} \end{bmatrix} = \begin{bmatrix} 0.00 & -0.127 & 0.00 \\ -0.675 & 0.00 & 0.00 \\ 0.00 & 0.00 & -0.594 \end{bmatrix}$$

$C_{z0}(\alpha, M_m) = C_{z1}(\alpha) + C_{z2}(\alpha)M_m, \quad C_{m0}(\alpha, M_m) = C_{m1}(\alpha) + C_{m2}(\alpha)M_m$

With:

$C_{z1}(\alpha) = -0.0015\alpha^3 + 0.0125\alpha^2 - 0.5052\alpha + 0.0429$

$C_{z2}(\alpha) = 0.0006\alpha^3 - 0.0138\alpha^2 + 0.1230\alpha - 0.0191$

$C_{m1}(\alpha) = -0.0055\alpha^3 + 0.2131\alpha^2 - 2.7419\alpha - 0.0381$

$C_{m2}(\alpha) = 0.0014\alpha^3 - 0.0623\alpha^2 + 0.8715\alpha - 0.4041$

$M_m$: is the March number defined as $M_m = V_m/c$

**Note:** this approximation is valid only for $0 \deg \leq \alpha \leq 25$ deg and $2.0 \leq M_m \leq 3.0$ which are satisfied in our case. ($\alpha = 10, M_m = 2.75$). The values of are given for an altitude of $40{,}000 ft$. The flight conditions are $V_m = 2662 ft/sec$ steady state angle of attack $\alpha_0 = 10°$

In the derivation of these equations it is assumed that the thrust forces $T_y, T_z$ components are negligible [29]. These equations will be taken as the state space representation of the missile autopilot, on which the control goals are specified. In the following section these equations will be linearized about an equilibrium point $(x_0 = (\alpha_0, 0,0,0,0,0))$ in order to get a LTI model. The linear dynamic model of BTT missile motion is given by the next state space equations:

$$\begin{bmatrix} \dot\alpha \\ \dot q \\ \dot\beta \\ \dot r \\ \dot p \\ \dot\phi \end{bmatrix} = \begin{bmatrix} Z_{\alpha\alpha} & 1 & Z_{\alpha\beta} & 0 & 0 & 0 \\ Z_{q\alpha} & 0 & Z_{q\beta} & 0 & 0 & 0 \\ Z_{\beta\alpha} & 0 & Z_{\beta\beta} & -1 & \alpha_0 & \frac{g}{V_m} \\ Z_{r\alpha} & 0 & Z_{r\beta} & 0 & 0 & 0 \\ Z_{p\alpha} & 0 & Z_{p\beta} & 0 & 0 & 0 \\ 0 & 0 & 0 & 0 & 1 & 0 \end{bmatrix} \begin{bmatrix} \alpha \\ q \\ \beta \\ r \\ p \\ \phi \end{bmatrix} + \begin{bmatrix} K_f C_{ze} & K_f C_{zr} & K_f C_{za} \\ K_{my} C_{me} & K_{my} C_{mr} & K_{my} C_{ma} \\ K_f C_{ye} & K_f C_{yr} & K_f C_{ya} \\ K_{mz} C_{ne} & K_{mz} C_{nr} & K_{mz} C_{na} \\ K_{mx} C_{le} & K_{mx} C_{lr} & K_{mx} C_{la} \\ 0 & 0 & 0 \end{bmatrix} \begin{bmatrix} \delta_e \\ \delta_r \\ \delta_a \end{bmatrix} \quad (30)$$

Where the $Z$ terms are constant coefficients called the flight derivatives given by:

$$\begin{cases} Z_{\alpha\alpha} = K_f \left.\frac{\partial C_{z0}}{\partial \alpha}\right|_{x_0}, & Z_{\alpha\beta} = K_f \left.\frac{\partial C_{z0}}{\partial \beta}\right|_{x_0} \\ Z_{q\alpha} = K_f \left.\frac{\partial C_{y0}}{\partial \alpha}\right|_{x_0}, & Z_{q\beta} = K_f \left.\frac{\partial C_{y0}}{\partial \beta}\right|_{x_0} \\ Z_{\beta\alpha} = K_{mx} \left.\frac{\partial C_{l0}}{\partial \alpha}\right|_{x_0}, & Z_{\beta\beta} = K_{mx} \left.\frac{\partial C_{l0}}{\partial \beta}\right|_{x_0} \\ Z_{r\alpha} = K_{my} \left.\frac{\partial C_{m0}}{\partial \alpha}\right|_{x_0}, & Z_{r\beta} = K_{my} \left.\frac{\partial C_{m0}}{\partial \beta}\right|_{x_0} \\ Z_{p\alpha} = K_{mz} \left.\frac{\partial C_{n0}}{\partial \alpha}\right|_{x_0}, & Z_{p\beta} = K_{mz} \left.\frac{\partial C_{n0}}{\partial \beta}\right|_{x_0} \end{cases}$$

Numerical values for missile's model [29] are given next in the following table:

Table: 1 Model parameters for the **HAVE DASH II BTT** missile.

| The parameter | Notation | Numerical value |
|---|---|---|
| The gravity | $g$ | 32.174 ft/sec² |
| Missile's mass | $m$ | 9.89 slug |
| Geometry coefficients | $k_f$ | 0.1534 ft² |
| | $k_m$ | 0.0959 ft³ |
| Sound velocity | $c$ | 968 ft/sec |
| Atmospheric density | $\rho$ | $5.124 \times 10^{-4}$ |
| The thrust force | $T_x$ | 389 lb |
| The inertia matrix entries | $I_{xx}$ | 1.1913 slug |
| | $I_{xx}$ | 100.5139 slug |
| | $I_{xx}$ | 100.5749 slug |

Numerical evaluation of the tabulated values in the parametric model yield:

$$\begin{bmatrix}\dot{\alpha}\\\dot{q}\\\dot{\beta}\\\dot{r}\\\dot{p}\\\dot{\phi}\end{bmatrix}=\begin{bmatrix}-0.0037 & 1 & 0 & 0 & 0 & 0\\-1.1462 & 0 & 0 & 0 & 0 & 0\\0 & 0 & -0.0044 & -1 & 0.1745 & 0.0121\\0 & 0 & 0.2770 & 0 & 0 & 0\\0 & 0 & 33.9063 & 0 & 0 & 0\\0 & 0 & 0 & 0 & 1 & 0\end{bmatrix}\begin{bmatrix}\alpha\\q\\\beta\\r\\p\\\phi\end{bmatrix}+\begin{bmatrix}-0.0019 & 0 & 0\\-2.3384 & 0 & 0\\0 & 0.0017 & 0\\0 & -2.0219 & 0\\0 & 0 & -37.1216\\0 & 0 & 0\end{bmatrix}\begin{bmatrix}\delta_e\\\delta_r\\\delta_a\end{bmatrix}$$

Where: $y_1 = \alpha$: The angle of attack  $\quad y_2 = \beta$: Small sideslip angle
$\quad\quad\quad y_3 = \phi$: The bank angle  $\quad u_1 = \delta_e$: Elevator deflection
$\quad\quad\quad u_2 = \delta_r$: Rudder deflection  $\quad u_3 = \delta_a$: Ailerons deflection

### 4.2. Block Pole Placement Comparison Study

In this paper, the proposed control scheme consisting of state feedback gain and feedforward gain is applied to the HAVE DASH II BTT missile using different state feedback methods discussed before; the results are assessed and compared in order to choose the suitable method for the design case. The characteristics upon which the comparison is based are:

- ➢ Controller gain magnitude.
- ➢ Transient response characteristics.
- ➢ Robustness in term of stability, performance and tracking.

The dimension of the matrix $A$ is $6 \times 6$ and the number of inputs is 3. The rank of the matrix $\Omega_c = [B\ AB]$ is 6, and then the system is block controllable of index 2. Therefore we can convert the system into block controller form by the following transformation matrix $T_c$:

$$T_c = \begin{bmatrix}E\Omega_c^{-1}\\E\Omega_c^{-1}A\end{bmatrix}, \quad \Omega_c = [B\ AB]^{-1} \text{ and } E = [O_3\ I_3]$$

We obtain the following:

$$A_c = \begin{bmatrix}O_3 & I_3\\-A_{c2} & -A_{c1}\end{bmatrix} = \begin{bmatrix}0 & 0 & 0 & 1 & 0 & 0\\0 & 0 & 0 & 0 & 1 & 0\\0 & 0 & 0 & 0 & 0 & 1\\-1.1462 & 0.0000 & 0.0000 & -0.0037 & 0.0000 & 0.0000\\0.0000 & -0.2770 & 0.8873 & 0.0000 & -0.0044 & -0.2084\\0.0000 & -1.8468 & 5.9163 & 0.0000 & -0.0015 & 0.0000\end{bmatrix}$$

$$B_c = \begin{bmatrix}E\Omega_c^{-1}\\E\Omega_c^{-1}A\end{bmatrix}B = \begin{bmatrix}O_3\\I_3\end{bmatrix}$$

We want to design a state feedback using block pole placement for the following set of desired eigenvalues: $\lambda_{1,2,3,4,5,6} = \{-4.90 \pm 7.35i, -17.1, -5.25, -7.5, -10.9\}$

- **Right solvents in diagonal form (Case I)**

$$K_{FB} = \begin{bmatrix} -15.4640 & -5.2726 & -26.8577 & -3.5605 & -0.0000 & 4.6460 \\ -18.2492 & 3.4630 & 9.8098 & -5.0280 & 0.0056 & -1.6785 \\ 0 & 0 & -0.9134 & -0.0000 & -0.7543 & -5.0211 \end{bmatrix}$$

$$K_{FF} = \begin{bmatrix} 15.9689 & 26.8727 & -4.6890 \\ 18.2589 & -9.9680 & 1.7393 \\ 0.0000 & -0.0000 & 5.0211 \end{bmatrix}$$

- **Right solvents in controller form (Case II)**

$$K_{FB} = \begin{bmatrix} 745.2977 & -8.8717 & -171.8956 & -899.1852 & 0.0352 & 90.1430 \\ -0.0012 & 0.4276 & 0.3576 & 0.0025 & 0.0326 & -0.0863 \\ 2464.8502 & -3.7787 & 1188.3972 & 2693.9081 & -0.8410 & -394.5275 \end{bmatrix}$$

$$K_{FF} = \begin{bmatrix} -744.7829 & 168.1091 & -79.2751 \\ -0.0000 & -0.4946 & 0.0863 \\ 2464.8608 & -1177.9664 & 361.9679 \end{bmatrix}$$

- **Right solvents in observable form (Case III)**

$$K_{FB} = \begin{bmatrix} -2469.6252 & -445.9117 & -494.3040 & 2857.7694 & -63.5247 & -101.8413 \\ -115.2148 & -21.8781 & 63.2212 & 133.3742 & -3.1904 & -19.8310 \\ 786.8736 & 138.3562 & 368.2749 & -909.6800 & 19.5929 & -4.5283 \end{bmatrix}$$

$$K_{FF} = \begin{bmatrix} 2471.3579 & 506.3381 & 67.3011 \\ 115.2758 & -62.7965 & 18.2190 \\ -787.2592 & -373.0190 & 15.5231 \end{bmatrix}$$

### 4.2.1. Time specifications

The magnitude and response characteristics for the three forms of solvents are summarized in the following two tables:

Table: 1 Gains magnitude for the three cases.

| Controller gains | Form of solvents | The 1-norm | The 2-norm | The ∞-norm | Forbenius norm |
|---|---|---|---|---|---|
| $K_{FB}$ | diagonal | 37.5808 | 31.9807 | 55.8008 | 38.9680 |
| | controller | $3.5931 \times 10^3$ | $4.0327 \times 10^3$ | $6.7463 \times 10^3$ | $4.0376 \times 10^3$ |
| | observer | $3.9008 \times 10^3$ | $4.0389 \times 10^3$ | $6.4330 \times 10^3$ | $4.0449 \times 10^3$ |
| $K_{FF}$ | diagonal | 36.8406 | 31.6246 | 47.5305 | 38.2114 |
| | controller | $3.2096 \times 10^3$ | $2.8559 \times 10^3$ | $4.0048 \times 10^3$ | $2.8606 \times 10^3$ |
| | observer | $3.3739 \times 10^3$ | $2.6641 \times 10^3$ | $3.0450 \times 10^3$ | $2.6731 \times 10^3$ |

Table: 2 Time specifications (response characteristics).

| The Outputs | Form of solvents | Percent undershoot (%) | Percent overshoot (POS %) | Settling time $t_s(s)$ | Rise time $t_r(s)$ |
|---|---|---|---|---|---|
| $\alpha$ | diagonal | -- | 17.2261 | 0.6731 | 0.1809 |
| $\alpha$ | controller | 667.1376 | -- | 1.2615 | 0.4523 |
| $\alpha$ | observer | 142.0237 | $2.103 \times 10^3$ | 0.8736 | 0.0114 |
| $\beta$ | diagonal | $9.7154 \times 10^{-4}$ | 3.3590 | 0.7696 | 0.3283 |
| $\beta$ | controller | 395.6040 | $2.9686 \times 10^3$ | 0.8868 | 0.0089 |
| $\beta$ | observer | $3.3867 \times 10^3$ | -- | 1.2577 | 0.4229 |
| $\phi$ | diagonal | -- | -- | 0.4529 | 0.2565 |
| $\phi$ | controller | $2.3281 \times 10^3$ | $1.7755 \times 10^3$ | 0.8755 | 0.0029 |
| $\phi$ | observer | $1.9255 \times 10^4$ | -- | 1.2723 | 0.4234 |

### 4.2.2. Robustness analysis:

**Robust stability:** Robust stability is determined using the measures defined in the previous section. First we find the norms of the left and right eigenvectors associated to each eigenvalue. The norms of the matrices consisting of the left and right eigenvectors respectively are:

- **Solvents in diagonal form:** $\|V\|_2 = 1.4595$ and $\|T\|_2 = \|V^{-1}\|_2 = 42.8363$
Hence the sensitivity of all eigenvalues is $s(V) = \|V\|_2 \|T\|_2 = 62.5211$

- **Solvents in controller form:** $\|V\|_2 = 1.9663$ and $\|T\|_2 = \|V^{-1}\|_2 = 1.5458 \times 10^4$
Hence the sensitivity of all eigenvalues is $s(V) = \|V\|_2 \|T\|_2 = 3.0394 \times 10^4$

- **Solvents in observer form:** $\|V\|_2 = 1.8185$ and $\|T\|_2 = \|V^{-1}\|_2 = 5.8622 \times 10^4$
Hence the sensitivity of all eigenvalues is $s(V) = \|V\|_2 \|T\|_2 = 1.0661 \times 10^5$

In the following the individual eigenvalues sensitivities are given for each of the three solvents from mentioned above.

Table: 3 Eigenvalues sensitivities.

| sensitivities | Diagonal solvents | Controller solvents | Observer solvents |
|---|---|---|---|
| $s(\lambda_{1,2} = -4.9 \pm 7.35i)$ | 8.2812 | $6.1437 \times 10^3$ | 384.4353 |
| $s(\lambda_3 = -17.1)$ | 30.2423 | $1.2298 \times 10^4$ | 781.5114 |
| $s(\lambda_4 = -5.25)$ | 7.5088 | $1.3700 \times 10^3$ | $2.0139 \times 10^4$ |
| $s(\lambda_5 = -7.5)$ | 8.6750 | $2.8577 \times 10^3$ | $4.7608 \times 10^4$ |
| $s(\lambda_6 = -10.9)$ | 30.2446 | $1.5880 \times 10^3$ | $2.7632 \times 10^4$ |

Table: 4 Stability measures.

| Stability measures | Diagonal solvents | Controller solvents | Observer solvents |
|---|---|---|---|
| $M_1 = \min_{\omega}\{\underline{\sigma}(A - BK - j\omega I)\}$ | 0.7226 | $1.3147 \times 10^{-3}$ | $3.2389 \times 10^{-3}$ |
| $M_2 = (s(\Lambda))^{-1}\|Re\{\lambda_n\}\|$ | 0.0784 | $1.6121 \times 10^{-4}$ | $4.5963 \times 10^{-5}$ |
| $M_3 = \min_{1 \leq i \leq 6}\{(s(\lambda_i))^{-1}\|Re\{\lambda_i\}\|\}$ | 0.2480 | $3.9842 \times 10^{-4}$ | $1.5753 \times 10^{-4}$ |

**Robust performance analysis:** We generate a random small perturbation using MATLAB software, we get the following:

$$\Delta A = \begin{bmatrix} 0.0069 & 0.0251 & 0.0047 & 0.0531 & 0.0100 & 0.0093 \\ 0.0256 & 0.0090 & 0.0046 & 0.0197 & 0.0038 & 0.0403 \\ 0.0195 & 0.0186 & 0.0084 & 0.0010 & 0.0074 & 0.0084 \\ 0.0164 & 0.0108 & 0.0142 & 0.0242 & 0.0184 & 0.0258 \\ 0.0072 & 0.0184 & 0.0043 & 0.0138 & 0.0204 & 0.0247 \\ 0.0165 & 0.0106 & 0.0119 & 0.0123 & 0.0273 & 0.0256 \end{bmatrix}, \quad \|\Delta A\|_2 = 0.1021$$

The new eigenvalues and corresponding change of each one are tabulated in table 5.

Table: 5 Change in eigenvalues due to random perturbation.

| Old eigenvalues | New eigenvalues | | | Relative change ($r_i$) | | |
|---|---|---|---|---|---|---|
| | Diagonal | Controller | Observer | Diagonal | Controller | Observer |
| $-4.9 + 7.35i$ | $-4.9937 + 7.5839i$ | $-15.1352$ | $-17.7855$ | 0.0285 | 2.4159 | 1.6793 |
| $-4.9 - 7.35i$ | $-4.9937 - 7.5839i$ | $-0.5110$ | $-4.0100 + 6.2699i$ | 0.0285 | 0.9691 | 1.5451 |
| $-17.1$ | $-16.1370$ | $-41.5796$ | $-11.3004 + 19.0094i$ | 0.0563 | 1.4316 | 1.1622 |
| $-5.25$ | $-5.2876$ | $-5.4459 + 2.0892i$ | $-2.0493$ | 0.0072 | 0.3997 | 0.6097 |
| $-7.5$ | $-7.2327$ | $-5.4459 - 2.0892i$ | $-4.0100 - 6.2699i$ | 0.0356 | 0.3906 | 0.9568 |
| $-10.9$ | $-11.8109$ | $-12.6083$ | $-11.3004 - 19.0094i$ | 0.0836 | 0.1567 | 1.7444 |

The tracking error due to this perturbation is:

- **Case of diagonal solvents:**
$$\Delta\alpha = 0.0024, \quad \Delta\beta = 0.0089, \quad \Delta\phi = 0.0082$$
- **Case of controller solvents:**
$$\Delta\alpha = -6.1205 \times 10^{17}, \quad \Delta\beta = 6.2267 \times 10^{18}, \quad \Delta\phi = 3.5858 \times 10^{19}$$
- **Case of observer solvents:**
$$\Delta\alpha = 0.0872, \quad \Delta\beta = -01952, \quad \Delta\phi = -1.1300$$

## 5. DISCUSSION OF THE RESULTS

Small gains are desirable because they minimize the control energy and prevent saturation of the controller elements and noise amplification. For time specifications, the smaller the settling time and maximum peak the better the time response. For the sensitivities of the eigenvalues, we choose the one that has the lowest sensitivity taking into account the distance of the eigenvalue from the $j\omega$ axis. For the robust stability the greater the value of its measure the more robustly stable the system, where $M_3$ is more accurate than $M_1$ and $M_2$ [1]. For robust performance, the smaller the value of relative change the better the performance. In our case, the crucial criterion is the robustness, because of the linearization of the model. From this results it can be observed that the diagonal solvents gives a smaller gains magnitude, better response, smaller sensitivities, and relatively higher performance measure. For the assumed perturbation, diagonal solvents gave less change in eigenvalues and less tracking errors.

It should be noted that for observer solvents presents highly undesired results, besides, for the considered perturbation the controller structure of the solvents resulted in unstable modes in the system.

## 6. CONCLUSION

The goal of the present work in this paper is to design 2DOF control law consisting of feedback gain and feedforward gain for the BTT missile autopilot. Given that the system is multivariable, the feedback gain that assign the eigenvalues to desired locations is not unique, different possibilities has been tested in this work. Of prime importance is that while the overall speed of response of the closed loop system is determined by its eigenvalues, the shape of the transient response is determined to a large extent by the eigenvectors, this is seen from the fact that while the settling time is around one second for the different cases, the response shapes are totally different.

The use of the three block canonical forms has shown that diagonal structure of solvents yield better results in term of gains, magnitudes, response and robustness, furthermore, the diagonal form of solvents is superior and less complexity in computations. In high performance missile design faster responses are needed, control efforts must be minimized, high oscillations are not tolerated and of course robustness is a must. According to the above discussion the block state feedback with diagonal solvents is the method to choose in this case of design.